\documentclass[12pt,preprint]{aastex}
\usepackage{graphicx}
\usepackage{latexsym,amssymb}
\usepackage{natbib}

\newcommand{\du}{\mathrm{d}}
\newcommand{\Mbh}{M_{\textrm{BH}}}
\newcommand{\Msph}{M_{\textrm{sph}}}
\newcommand{\Mgal}{M_{\textrm{gal}}}

\newcommand{\Mnuc}{M_{\textrm{NC}}}

\newcommand{\Lgal}{L_{\textrm{gal}}}
\newcommand{\Lnuc}{L_{\textrm{NC}}}

\newcommand{\MLgal}{\Upsilon_{\textrm{gal}}}
\newcommand{\MLnuc}{\Upsilon_{\textrm{NC}}}
\newcommand{\MLcmo}{\Upsilon_{\textrm{CMO}}}

\begin{document}

\title{Formation of central massive objects via tidal compression}

\author{Eric Emsellem}
\affil{Universit\'e de Lyon, Lyon, F-69003, France ; Universit\'e Lyon~1,
Observatoire de Lyon, 9 avenue Charles Andr\'e, Saint-Genis
Laval, F-69230, France ; CNRS, UMR 5574, Centre de Recherche Astrophysique
de Lyon ; Ecole Normale Sup\'erieure de Lyon, Lyon, F-69007, France}
\email{emsellem@obs.univ-lyon1.fr}

\and

\author{Glenn van de Ven\altaffilmark{1}}
\affil{Institute for Advanced Study, Einstein Drive, Princeton, NJ
  08540, USA} 
\email{glenn@ias.edu}

\altaffiltext{1}{Hubble Fellow}

\begin{abstract}
For a density that is not too sharply peaked towards the center, the
local tidal field becomes compressive in all three directions.
Available gas can then collapse and form a cluster of stars in the
center, including or even being dominated by a central black hole.
We show that for a wide range of (deprojected) S\'ersic profiles in
a spherical potential, the tidal forces are compressive within a
region which encloses most of the corresponding light of observed
nuclear clusters in both late-type and early-type galaxies.  In such
models, tidal forces become disruptive nearly everywhere for
relatively large S\'ersic indices $n \ga 3.5$.  We also show that
the mass of a central massive object (CMO) required to remove all
radial compressive tidal forces scales linearly with the mass of the
host galaxy. If CMOs formed in (progenitor) galaxies with $n \sim
1$, we predict a mass fraction of $\sim 0.1-0.5$\,\%, consistent
with observations of nuclear clusters and super-massive black holes.
While we find that tidal compression possibly drives the formation
of CMOs in galaxies, beyond the central regions and on larger scales
in clusters disruptive tidal forces might contribute to prevent gas
from cooling.
\end{abstract}


%============================= section 1 =============================
\section{Introduction}
\label{sec:intro}
%=====================================================================

It is now well-known that the masses of supermassive black holes
(SMBHs) in the centres of galaxies and bulges correlate with the
stellar velocity dispersion, $\Mbh \propto \sigma_\star^{\alpha}$ with
$\alpha \sim 4-5$ \citep[e.g.][]{Ferrarese+01, Gebhardt+01,
  Tremaine+02}, as well as nearly linearly with the mass of these
spheroids, $\Mbh \propto \Msph^{1.12\,\pm\,0.06}$ \citep[e.g.][]{MF01,
  HaringRix04}. \cite{Ferrarese+06}, \cite{WH06} and \cite{Rossa+06}
also found that the masses of nuclear (star) clusters (NCs), which are
present in many both late and early-type galaxies \citep[see,
e.g.][]{Boeker+02, Cote+06}, are similarly related to the properties
of the host galaxy \citep[see also][]{GD07}. Recently,
\cite{McLaughlin+06} proposed momentum feedback, from accretion onto
SMBHs or from stellar and supernovae winds in the case of NCs, as an
explanation for the observed relations.

We investigate if central massive objects (CMOs), in the form of NCs,
may have formed from gas being tidally compressed in the centers of
galaxies. This effect, resembling compressive shocking of globular
clusters by the Galactic disk \citep{Ostriker+72}, has been studied by
\cite{Valluri93} in the context of tidal compression of a (disk)
galaxy in the core of galaxy cluster. At the scale of galaxies,
\cite{DasJog99} argue for tidal compression of molecular clouds in the
centers of flat-core early-type galaxies and ultraluminous galaxies as
an explanation for the presence of observed dense gas. Very recently,
an independent study by \cite{Masi07} emphasised the potential
importance of compressive tidal forces.

Low luminosity early-type galaxies and late-type galaxies share an
overall luminosity profile with relatively low central power-law
slopes. The fact that NCs are predominantly found in such galaxies
\citep[see e.g.][]{Cote+06} may provide an interesting link between
the presence of CMOs in galaxies and the properties of the host
galaxy. In the present study, we investigate whether tidal forces may
help explaining this link. We first derive the radial component of the
tidal force associated with a (deprojected) S\'ersic profile in
\S~\ref{sec:tidalcompression}. We then examine in \S~\ref{sec:scaling}
how this applies to simple models of CMO hosts, including early-type
and late-type galaxies. The corresponding results are then briefly
discussed in \S~\ref{sec:disc}, and conclusions are drawn in
\S~\ref{sec:concl}.

%============================= section 2 =============================
\section{Tidal compression}
\label{sec:tidalcompression}
%=====================================================================

The radial component of the tidal field in a spherical potential is
given by
\begin{equation}
  \label{eq:Trad}
  T_R(r) = 4 \pi G \left[ 2\langle\rho\rangle/3 - \rho(r) \right] R,
\end{equation}
at a distance $R$ from the position at radius $r$ about which the
gravitational field is expanded to first order. This radial component
is compressive ($T_R<0$) if the local density $\rho(r)$ is larger than
$2/3$ of the mean density $\langle\rho\rangle = 3 M(<r)/(4 \pi r^3)$
within the radius $r$. The two components perpendicular to the radial
direction are always compressive, but with varying magnitude, so that
the tidal field is generally anisotropic.
%% \citep[see e.g.][and references therein]{Valluri93, DasJog99, Masi07}. 
However, since the compression of a gas cloud tends to become rapidly
isotropic, we only consider the radial component \citep[see
also][]{DasJog99}.

%============================= subsection 2.1 =============================
\subsection{Density profiles}
\label{sec:sersic}
%==========================================================================

It is today well-known that the surface brightness profiles of
early-type galaxies as well as bulges are overall well fitted by a
S\'ersic (\citeyear{Sersic63, Sersic68}) profile
\begin{equation}
  \label{eq:rhoSersic}
  I(R) = I_e \exp\left\{
      -b_n \left[ \left(\frac{R}{R_e}\right)^{1/n} - 1 \right] \right\},
\end{equation}
with $I_e$ the surface brightness at the half-light radius $R_e$, and
$b$ a constant that depends on the index $n$. The latter follows by
solving $\Gamma(2n) = 2 \Gamma(2n,b)$, with $\Gamma$ the gamma
function, but to high precision can be approximated by $b_n = 2\,n -
1/3 + 4/405\,n + 46/25515\,n^2$ \citep{1999A&A...352..447C}.

The deprojection of the S\'ersic profile (assuming spherical symmetry)
has to be done numerically \citep{1991A&A...249...99C}.
However, the analytic density profile of \cite{PrugnielSimien97}
\begin{eqnarray}
  \label{eq:rhoPS}
  \rho^{PS97}(r) = \rho_e \left(\frac{r}{R_e}\right)^{-p_n}
  \exp\left\{
    -b_n\left[ \left(\frac{r}{R_e}\right)^{1/n} - 1 \right] \right\}.
\end{eqnarray}
provides in projection a good match\footnote{Although not significant,
  please note that there is a typo in the expression of $p$ in most
  published papers (with $0.05563/n^2$ instead of $0.05463/n^2$), the
  best fit as given in \cite{LimaNeto+99} being the one mentioned
  here.} to the S\'ersic profile when $p_n=1.0-0.6097/n+0.05463/n^2$
and a constant (stellar) mass-to-light ratio is assumed. When $p_n=0$,
the density profile reduces to the `intrinsic' S\'ersic profile, also
known as Einasto's model \citep[see][and references therein]{Einasto+89}. Recently, it has been
shown that both the Prugniel-Simien and Einasto models provide a very
good fit to simulated dark matter halos, better than a (generalized)
NFW profile \citep[e.g.][]{Merritt+06}.

The Prugniel-Simien profile is a reasonable approximation for the
deprojected S\'ersic profile for a relatively wide range of values of
$n$ \citep{Marquez+00}, but not as accurate as originally claimed.
\cite{Trujillo+02} proposed a much improved expression using modified
Bessel functions, with relative errors less than $0.1$\,\% in the
radial range $10^{-3} \le r/R_e \le 10^3$ for $n>1$. We extended their
expression to
\begin{equation}
  \label{eq:Abel}
  \rho(r) =  \Upsilon \, \frac{I_e \exp(b_n) b_n}{n \pi R_e} \, 
  \frac{(r/R_e)^{k(1-n)/n} \, 2^{(n-1)/2n} \, K_\nu\left[b_n(r/R_e)^{1/n}\right]}
  {1- \Sigma_{i=0}^m \, a_i \, \log(r/R_e)^i},
\end{equation}
where $\Upsilon$ is the (constant) mass-to-light ratio and $K_\nu[x]$
is the $\nu$\,th-order modified Bessel function of the third kind. The
fitting parameters are $\nu$, $k$ and the coefficients $a_i$
($i=0,1,\dots,m$) of the $m$\,th-order polynomial in $\log(r/R_e)$.
With a cubic polynomial ($m=3$), this approximation results in a fit
with residuals still of the order of $0.1$\,\% or (often
significantly) less, but this time for all values between $0.5 \le n
\le 10$ (see Appendix~\ref{app:Abel} for details). It surpasses (by a
factor $>100$) the Prugniel-Simien profile.

%%FIG
\begin{figure}
  \begin{center}
    \includegraphics[width=1.0\textwidth,trim=0 0 0 0 mm]{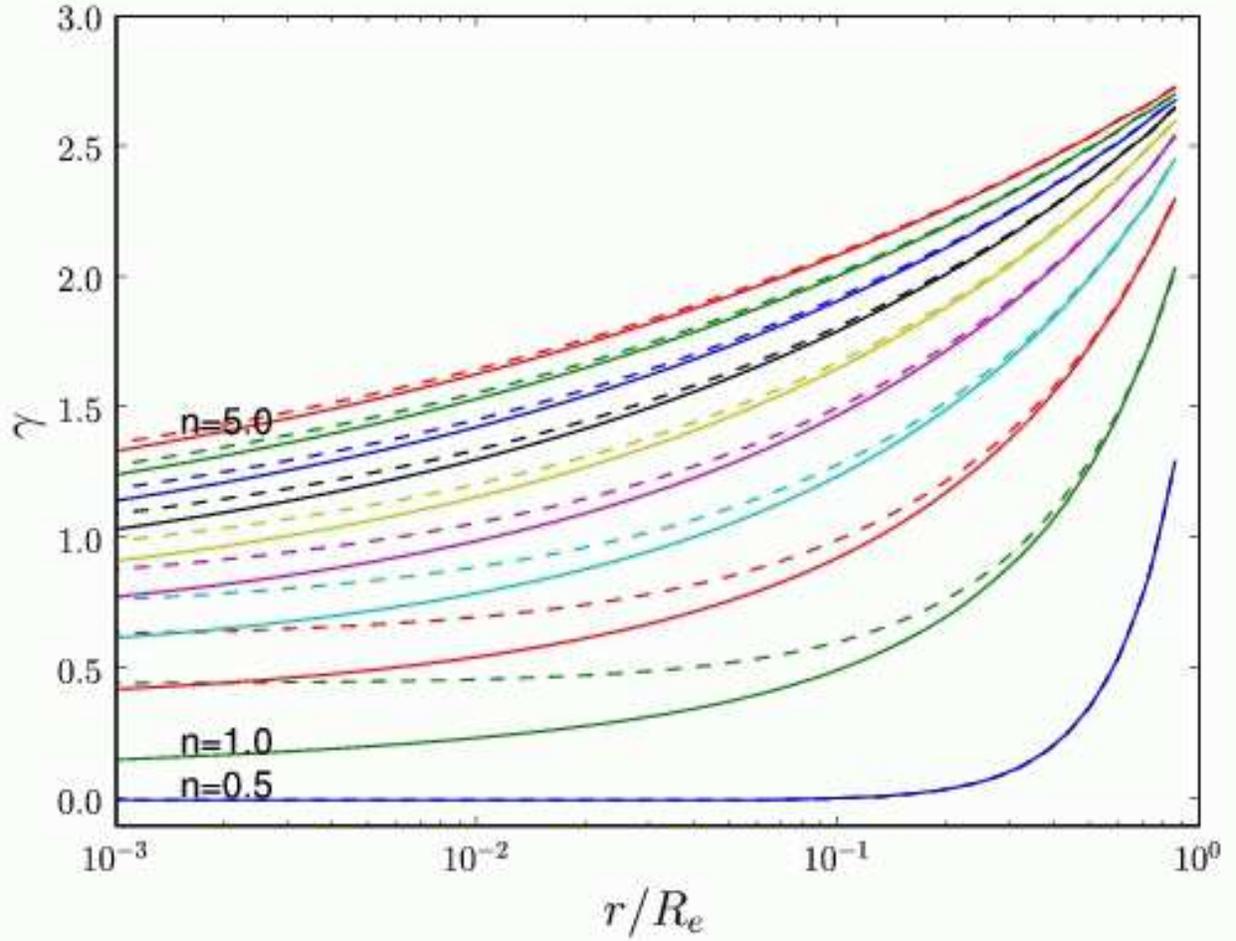}
  \end{center}
  \caption{Slopes $\gamma$ of the intrinsic density associated with the deprojected
    S\'ersic profile (solid lines) and the Prugniel-Simien model with
    $p_n=1.0-0.6097/n+0.05463/n^2$ (dashed lines), for values of $n$
    from 0.5 to 5 (with steps of 0.5).}
  \label{fig:slopes}
\end{figure}
%%FIG

In the following, we favour Eq.~\ref{eq:Abel} as approximation for the
intrinsic density $\rho(r)$, but also consider the Prugniel-Simien
profile in Eq.~\ref{eq:rhoPS}. To compute $\langle \rho \rangle$ in
Eq.~\ref{eq:Trad} for the radial component of the tidal field, we also
need the enclosed mass $M(r) \equiv 4\pi \int_0^r \rho(r') {r'}^2
\du{r'}$, which in case of the Prugniel-Simien profile reduces to
$M^{PS97}(r) = 4 \pi R_e^3 \rho_e \exp(b_n) n b_n^{(p_n-3)n}
\gamma[(3-p_n)n,b_n(r/R_e)^{1/n}]$, with $\gamma[a;x]$ the incomplete
gamma function. The latter should not be confused with the (negative)
slope of the density $\gamma(r) \equiv -\du\log\rho/\du\log r$, which
for the Prugniel-Simien model becomes $\gamma^{PS97}(r) = p_n + b_n
(r/R_e)^{1/n}/n$. In Fig.~\ref{fig:slopes}, we present $\gamma(r)$ for
both cases out to the effective radius. For large values of $n$ ($\ga
5$), there is no significant difference in the values of $\gamma$ for
$10^{-3} < r / R_e < 1$, but for lower values of $n$, the
Prugniel-Simien profile tends to have slightly larger $\gamma$ slopes
at small radii.

%======================= subsection 2.2 =============================
\subsection{Tidal field}
\label{sec:tidalfield}
%=====================================================================

For a spherical model with an average density $\overline{\rho}(r)
\propto r^{- \gamma}$, we have $\rho(r) / \overline{\rho}(r) = 1 -
\gamma / 3$ and the radial component of the tidal field is then
proportional to $(\gamma - 1) \overline{\rho}(r) R$: it is therefore
negative (compressive) only when $\gamma \le 1$. This becomes less
straightforward when $\gamma$ varies with radius, but
Fig.~\ref{fig:slopes} still suggests that we should expect compressive
tidal forces for S\'ersic indices $n \la 3.5$ at least within $r/R_e
\la 10^{-3}$. For a typical galaxy, the latter corresponds to a few pc
and is similar to the smallest observed half-light radii of NCs (see
also Fig.~\ref{fig:rh_n}).

%%FIG
\begin{figure}
  \begin{center}
    \includegraphics[width=1.0\textwidth,trim=0 0 0 0 mm]{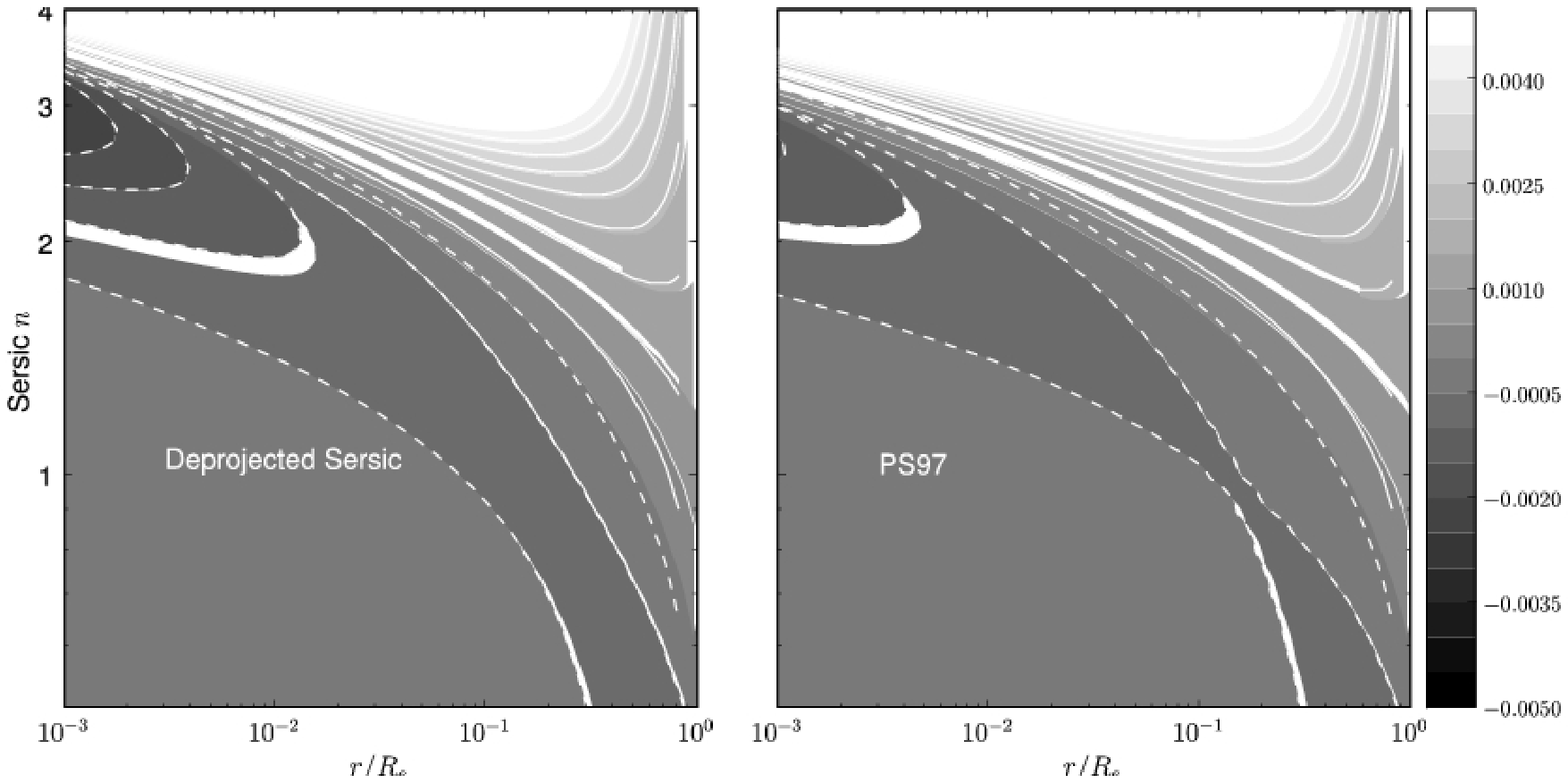}
  \end{center}
  \caption{Radial component of the tidal field (per unit mass) in a spherical
    potential for the deprojected S\'ersic profile (left) and the
    Prugniel-Simien model (right) for different values of $n$ and for
    radii between $10^{-3}$ and $1$ $R_e$. The dashed lines show where
    the radial tidal field is compressive.}
    %% For this plot, the total luminosity $L_\mathrm{tot}$ of each
    %% model has been set to 1. 
  \label{fig:Trad}
\end{figure}
%%FIG

This is confirmed in Fig.~\ref{fig:Trad}, where we show the radial
component of the tidal field (per unit mass) for different values of
the S\'ersic index $n$. Both the deprojected S\'ersic approximation in
Eq.~\ref{eq:Abel} (left panel) and the Prugniel-Simien profile in
Eq.~\ref{eq:rhoPS} (right panel) yield similar behaviour: $T_R < 0$ is
found only for $n \la 3.5$ given the same minimum radius of $10^{-3}
R_e$ as above, and $r/R_e \la 1$ given as minimum S\'ersic index that
of an exponential disk ($n=1$). Between these limits, there is a range in
radii from the center to a truncation radius $r_t$, within which the
tidal field is compressive. Within this range, the radius $r_{minT_R}$
at which the radial force reaches its minimum, increases with
decreasing $n$ index, with a reasonably good approximation given by $n
= -0.73 \, \log{(r_{minT_R}/R_e) + 0.38}$.

For the Prugniel-Simien profile, $n \la 3.5$ corresponds to $p_n \la
0.83$. As expected, forcing the latter inner slope to small values
(e.g., $p_n=0$) implies that the tidal forces are compressive for
rather large $n$ and for a larger radial range at a given $n$.  For
$p_n = 1$, the radial tidal force per unit mass is obviously positive
everywhere.

%============================= section 3 =============================
\section{Scaling relations}
\label{sec:scaling}
%=====================================================================
 
The above analysis shows that the inner slope $\gamma$, which is a
function of the S\'ersic index $n$, determines the truncation radius
$r_t/R_e$ within which there exists compressive tidal forces. If we
now assume that these negative tidal forces play a role to trigger
star formation from passing-by molecular clouds, we can expect that
$r_t$ provides a scaling for the size of the CMO as function of $n$
and $R_e$.

%======================= subsection 3.1 =============================
\subsection{Truncation radius}
\label{sec:truncationradius}
%=====================================================================

\begin{figure}
  \begin{center}
    \includegraphics[width=1.0\textwidth,trim=0 0 0 0 mm]{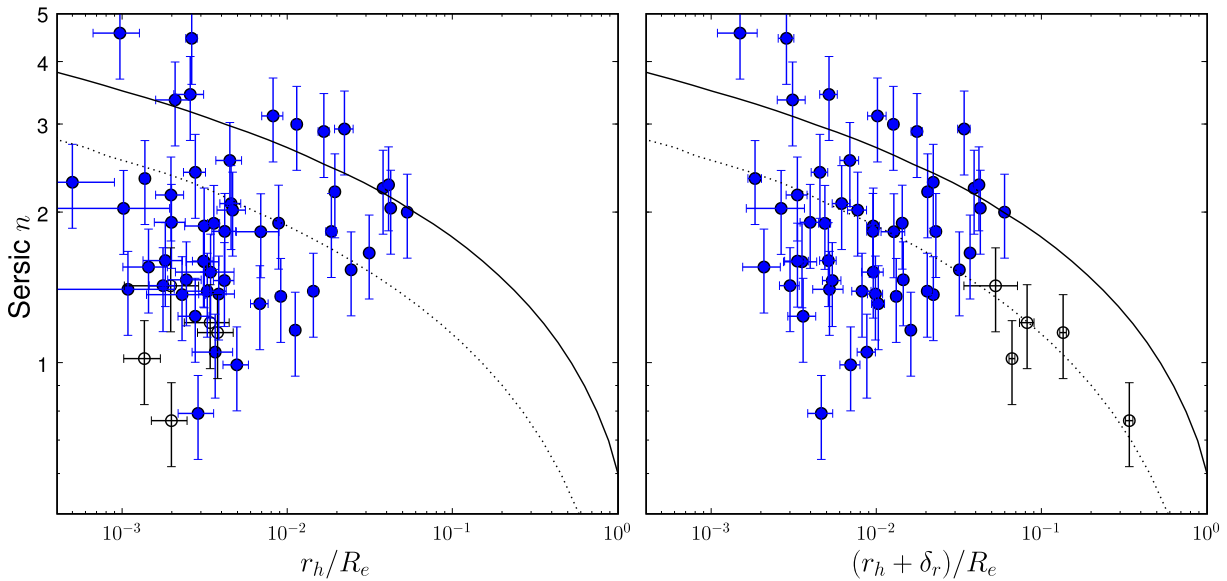}
  \end{center}
  \caption{S\'ersic index $n$ versus the half-light radius $r_h$ (normalised by
    the effective radius $R_e$ of the host galaxy) for the NCs
    detected by \cite{Ferrarese+06b} and \cite{Cote+06} (circles with
    error bars). Uncertainties on $n$ and $r_h$ have been assumed to
    be 10\,\% and 0.01 arcsec, respectively  \citep[see e.g., Fig.~114 of][]{Ferrarese+06b}.  The empty circles
    correspond to the 5 offset NCs revealed by \cite{Ferrarese+06b}
    and \cite{Cote+06}.  In the right panel, the measured offsets
    $\delta_r$ have been added to the half-light radius $r_h$. The
    solid line corresponds to the truncation radius, at each given
    $n$, within which the tidal forces are compressive for a
    deprojected S\'ersic profile, and the dotted line indicates the
    location of the minimum (negative) radial tidal force (see also
    Fig.~\ref{fig:Trad}).}
  \label{fig:rh_n}
\end{figure}

For 51 early-type galaxies in the Virgo cluster, \cite{Ferrarese+06b}
and \cite{Cote+06} have detected a central NC, for which they measured
the half-light radius $r_h$, as well as the S\'ersic parameters $n$
and $R_e$ of their host galaxy. They have also included the
characteristics of 5 NCs which were detected but offset from the
photometric centre: in these five cases, the bright NCs were located
within about 5\arcsec\ from the centre. In the left panel of
Fig.~\ref{fig:rh_n}, we show $n$ versus $r_h/R_e$ of the observed NCs,
as compared to the truncation radius $r_t/R_e$ for the deprojected
S\'ersic profile. In the right panel, we also take into account the
offsets $\delta_r$ measured by \cite{Cote+06}, which clearly
emphasises the 5 special cases mentioned above (for the other NCs, the
addition of $\delta_r$ does not significantly change their position in
the diagram). 

Only two NCs (in NGC\,4578 and NGC\,4612) are significantly outside
the presumed region where compressive tidal forces exist. Remarkably,
the five offset NCs lie within that region too, rather close to the
radius of minimum (negative) compressive tidal force, when including
their apparent position with respect to the galaxy centre. This
indicates that the stars in these clusters may have formed in a region
where the tidal forces were negative, or moved via dynamical friction
within that region since their formation with typical timescales of
the order of a few hundreds Myr. Moreover, as we argue below,
especially the old NCs in early-type galaxies might have already
formed in their possibly gas-rich and spiral-like progenitors with
S\'ersic index values lower than the currently measured $n$ values. As
a result, this can move the points in Fig.~\ref{fig:rh_n} downwards
more into the region of compressive tidal forces, and, at the same
time, the variation in progenitors as well as merging history might
contribute to the large scatter observed.

%%FIG
%%FIG

%======================= subsection 3.2 =============================
\subsection{Central mass}
\label{sec:entralmass}
%=====================================================================

The presence of an additional (compact) CMO in a galaxy increases the
averaged mass density $\langle\rho\rangle$ (see Eq.~\ref{eq:Trad}),
and therefore directly reduces the size of the radial region where
tidal forces are compressive. Henceforth, given a host galaxy with
S\'ersic parameters $n$, $R_e$, and $I_e$ (or total luminosity $\Lgal$)
and mass-to-light ratio $\MLgal$, there is therefore a mass $M_+$
above which the radial tidal force $T_R$ becomes positive (disruptive)
everywhere. When adding a central mass intermediate between
0 and $M_+$, negative $T_R$ values are restricted to
a ring-like region. Offset stellar nuclei could form there, 
and subsequently take significant time before being dragged into the centre
via dynamical friction. For large S\'ersic index $n \ga 3.5$, $M_+ \approx 0$
since $T_R$ is already positive nearly everywhere. For $n \la 3.5$, we
(numerically) derive $M_+$ from the maximum of $r^3 \left[3 \rho(r) -
  2\langle\rho\rangle \right] / 3$, and corresponding luminosity $L_+$
for a given mass-to-light ratio $\MLcmo$ of the CMO. When $\MLcmo
\approx \MLgal$, the mass-to-light ratio cancels out and we obtain
directly $L_+$ for given $n$, $R_e$ and $\Lgal$ of the host galaxy.

%%%FIG
\begin{figure}
  \begin{center}
    \includegraphics[width=1.0\textwidth,trim=0 0 0 0 mm]{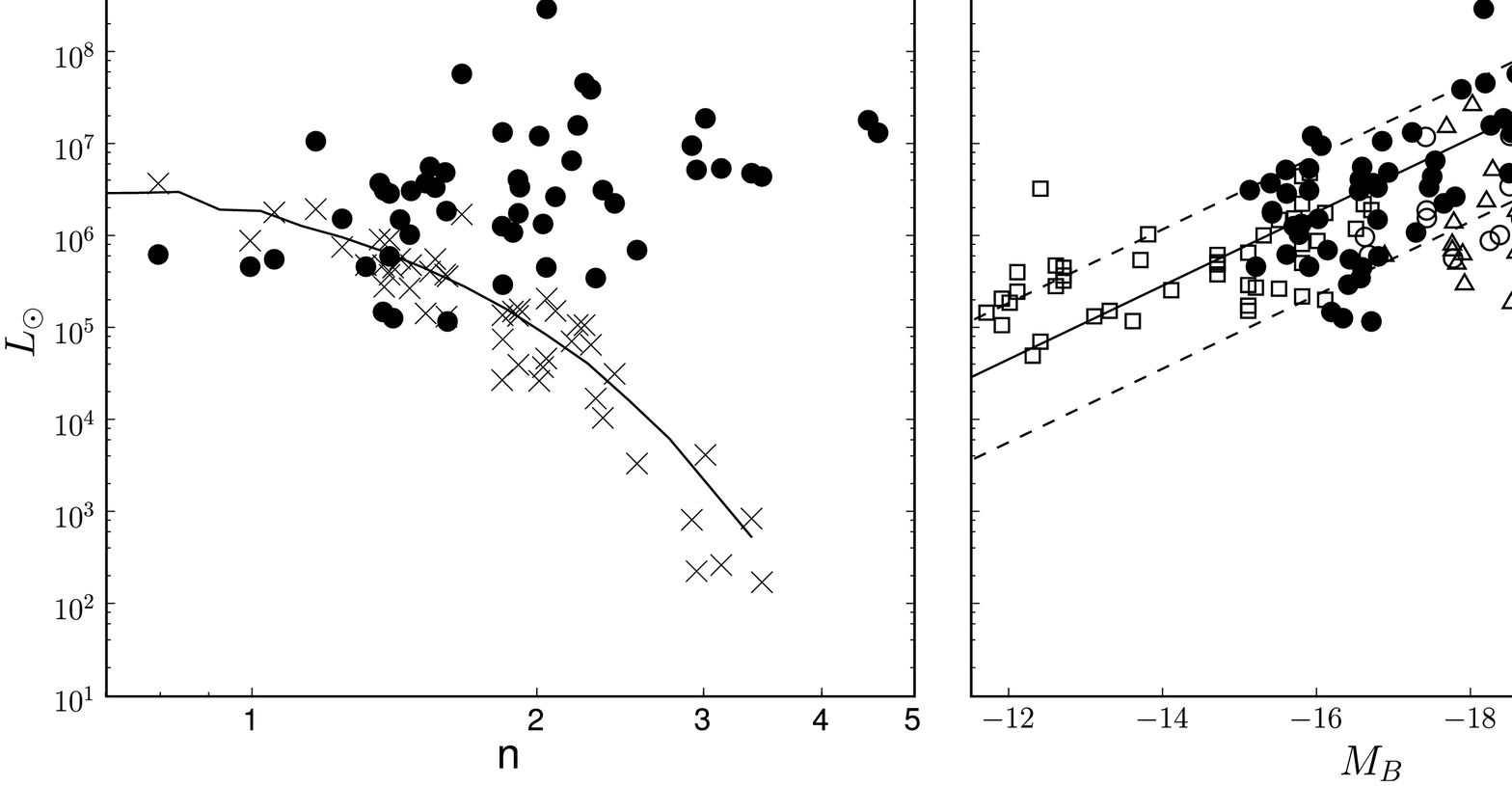}
  \end{center}
  \caption{Nuclear cluster luminosity $L_{NC}$ (black circles) versus
    the S\'ersic index $n$ (left panel) and the absolute $B$ band
    magnitude of the host galaxy (right panel). The crosses in the left panel
    represent the predictions for $L_+$ for the Virgo
    NCs observed by \citep{Ferrarese+06b}, using the measured 
    $n$ of the host galaxy. 
    The black solid line shows $L_+$ as derived from the scaling laws described in
    the main text. In the right panel, empty circles correspond to
    the NCs observed in late-type spirals \citep{Rossa+06}, empty
    triangles to NCs in early-type spirals \citep{Carollo+02}, empty
    squares to dwarf ellipticals \citep{Lotz+04}, and the black (solid
    and dashed) lines to $L_+$ prediction for models where the
    S\'ersic index $n$ was fixed to three different values (0.5, 1,
    1.5 as indicated).  The observed $B$ band luminosities for the NCs
    observed by \cite{Carollo+02} and \cite{Lotz+04} were converted
    from the $V$ band, using a mean $B-V = 1$.}
  \label{fig:Lnuc}
\end{figure}
%%%FIG

Adopting the latter assumption, we compute $L_+$ corresponding to the
measured parameters $(M_B, n, R_e)$ for each of the Virgo early-type
galaxies studied by \cite{Ferrarese+06b} and \cite{Cote+06}. In
Fig.~\ref{fig:Lnuc}, we show the resulting $L_+$ values (crosses)
together with the measured luminosity $\Lnuc$ of the observed NCs
\citep{Cote+06} versus $n$ and $M_B$ of the host galaxy (left and
right panel, respectively). The solid line in the left panel shows the
prediction of $L_+$ as function of $n$ based on empirical scaling
relations of galaxies.  Here we use the global correlations derived in
\cite{Andredakis+95} and \cite{Graham01} to obtain $\log\,n = - 0.216
\, (M_B + 18) + 0.44$ and (for $R_e$ in kpc) 
$\log\,R_e = - 0.277 \, (M_B + 18) + 0.4$.
For a given ($B$-band) luminosity of the host galaxy these relations
provide us with a pair of $n$ and $R_e$, so that together they yield
an estimate of $L_+$.  Although these relations significantly differ
from $\log\,n = -0.10 \, (M_B + 18) + 0.39$ and $\log\,R_e = - 0.055
\, (M_B + 18) + 0.04$ derived by \cite{Ferrarese+06b} for Virgo
early-type galaxies, they are still consistent with the observed
measurements, and have the advantage to also fit the observed
correlation for spiral bulges.

For large values of $n$, the predicted $L_+$ are much smaller than the
observed CMO luminosities $L_{NC}$, as expected. Only for values of $n
\la 1.5$, the predicted $L_+$ are in the same luminosity range as the
corresponding observed $L_{NC}$ values.  In the right panel, we have
added samples of NCs observed in dwarf ellipticals \citep{Lotz+04}
(empty squares), in early-type spirals by \cite{Carollo+02} (empty
triangles) and in late-type spiral galaxies by \cite{Rossa+06} (empty
circles). Again we see that the predicted $L_+$ are much lower than
the observed $L_{NC}$ values in early-type galaxies, except for faint
galaxies with $M_B$ around $-16$. Although we do not know the S\'ersic
index associated with the spiral and dwarf elliptical hosts of
observed NCs, we can safely assume that it should be low with $n \sim
1$ being a good approximation for the overall surface brightness
profile. It is interesting to note that most spiral NCs lie {\em
  below} the $L_+$ curve with $n = 1$, and all of them are below the
one with $n=0.5$.

%============================= section 4 =============================
\section{Discussion}
\label{sec:disc}
%=====================================================================

The frequency of occurence of NCs in both intermediate luminosity
early-type galaxies, and in spiral galaxies is reported to be between
about 60 and 80\,\% \citep{Carollo+98, Carollo+02, Boeker+02, Cote+06,
  Balcells+07a}.  For nearly all galaxies in the samples mentioned
above, the observed nuclear clusters have individual luminosities
$L_{NC}$ in a range similar to (for early-type galaxies and dwarfs
ellipticals) or lower than (for spirals) the $L_+$ values derived from
the total galaxy luminosity, considering that each NC and its host
galaxy have the same mass-to-light ratio, and assuming a spherical
model with $n \sim 1$ (exponential profile).

Even though the density distribution of most galaxies is clearly not
spherical, the tidal forces are determined by their gravitational
potential, which in general is significantly rounder than the density,
even for spiral galaxies. For example, an axisymmetric logarithmic
potential, which is only about a third as flattened as the
corresponding density distribution \citep[e.g.\ \S~2.2.2 of][]{BT87},
lowers the predicted $L_+$ values by a factor of about two if the
flattening of the potential is reduced from unity (sphere) to one-half
(disk). The latter would then only shift down the lines at fixed $n$
shown in Fig.~\ref{fig:Lnuc} by $\sim 0.3$ dex.

The assumption of equal mass-to-light ratio $\MLnuc = \MLgal$ between
NC and host galaxy might have a stronger effect.  NCs are sometimes
relatively bluer than their host galaxies (e.g., in late-type spiral
galaxies), so that we can expect $\MLnuc < \MLgal$ \citep[by a factor
of 1 to 10, see e.g.][]{Boeker+04}. This implies that most of these
NCs have masses which would in fact lie even {\em lower}, with respect
to the predictions of $M_+$, than presently illustrated in
Fig.~\ref{fig:Lnuc}.  On the other hand, their host galaxies typically
have $n \approx 1$ and are gas-rich, so that one might expect ongoing
star formation driven by the significant tidal compression. Indeed,
late-type spiral galaxies seem to witness recurrent star formation in
their central regions and most of the NCs observed in these galaxies
correspond to multiple formation episodes. Henceforth, the NC might be
still building up its mass $\Mnuc$ towards the predicted $M_+$ value.
Moreover, the youngest population of stars likely dominates the total
light $\Lnuc$ of the cluster and hence the (measured) mass-to-light
ratio $\MLnuc$, but contributes (much) less to its mass $\Mnuc$
\citep[e.g.][]{Walcher+06, Rossa+06}. Finally, the seemingly strong
link between the luminosity (and mass) of the NC and its host galaxy
might only hold true for the centrally dominating spheroidal component
\citep[e.g.][]{Rossa+06, WH06}. Whereas in early-type galaxies the
spheroidal component dominates (the light of) the galaxy, the
bulge-to-total-light ratio decreases significantly towards later-type
galaxies, with $B/T \sim 1/4$ already for lenticular (S0) and
early-type spiral (Sa) galaxies \citep[e.g.][]{Balcells+07b} and $B/T
\la 1/10$ for later-type spiral galaxies \citep[e.g.][]{Graham01}. As
a result, if instead of the total luminosity, we compare in the right
panel of Fig.~\ref{fig:Lnuc} $\Lnuc$ with the spheroidal luminosity,
we expect the NCs in the spiral galaxies to move by a maximum of $\sim
2.5$ dex to the left, i.e., closer to the predicted line for $n=1$.
Also, this implies that nearly all NCs appear in spheroids fainter
than $M_B = -18$, consistent with the transition between NCs and SMBHs
as emphasised by \cite{Cote+06} and \cite{Ferrarese+06}.

Although $\MLnuc$ might be similar to the \emph{stellar} mass-to-light
ratio of the host galaxy (especially in early-type galaxies with old
stellar populations), its \emph{total} mass-to-light ratio $\MLgal$
might be different and vary among galaxies due to a contribution from
possible dark matter. The tilt of the Fundamental Plane of elliptical
galaxies implies $\MLgal \propto L^{0.3}$
\citep[e.g.][]{Jorgensen+96}, so that $\Lnuc \propto \Lgal^{1.3}$.
This results in a steeper slope in the right panel of
Fig.~\ref{fig:Lnuc} compared to the predicted relations (solid and
dashed lines), and similarly a shallower slope if $\MLgal$ decreases
with $\Lgal$. There is an indication that the NCs in observed
early-type galaxies (filled circles) indeed follow a slightly steeper
slope. The NCs in late-type spirals (open circles) and dwarf
ellipticals (open squares) seem to indicate a shallower slope,
consistent with the corresponding correlations $\Lnuc \propto
\Lgal^{0.78}$ and $\Lnuc \propto \Lgal^{0.87}$ derived by
\cite{Boeker+04} and \cite{GrahamGuzman03}, respectively. In all
cases, when converting from luminosities to masses the slope
approaches unity \citep[see also][]{Ferrarese+06, WH06}, consistently
with the predicted linear scaling of $M_+$ with $\Mgal$ at fixed $n$.

In other words, if we, from now on, assume that all these observed NCs
originated in host galaxies with nearly exponential surface brightness
profiles ($n\sim1$), they would have formed in the presence of
compressive tidal forces. The amplitude of these compressive tidal
forces directly scales with the total mass $\Mgal$ of the host galaxy
(or with the component which dominates the central mass profile). For
the models described above, $M_+$ thus linearly scales with the host
mass at fixed $n$. If compressive tidal forces dictate the efficiency
of the triggering of star formation in the central regions of such
galaxies, they would impose a direct linear scaling between the mass
of the formed NCs and the host mass. By changing the S\'ersic index
$n$ of the models, we obviously shift the relation between $M_+$ and
$\Mgal$, predicting lower $M_+$ for a larger $n$ at a given $\Mgal$.
For a deprojected S\'ersic profile, we find $\log{(M_+/\Mgal)} \sim
-1.9 \, n - 0.4$, which implies a ratio of $0.5$\,\% for $n=1$.
\cite{Cote+06} mention a typical ratio of nucleus-to-galaxy luminosity
of $0.3$\,\%, and \cite{Ferrarese+06} a corresponding mass ratio of
$0.18$\,\% for CMOs (NCs and SMBHs) in early-type galaxies. The
corresponding S\'ersic index $n$ are respectively $\sim 1.1$ and $\sim
1.2$, i.e., close to unity in both cases.

The picture in which NCs form in galaxy hosts with $n \sim 1$ seems
appropriate for spiral galaxies \citep[see][]{Seth+06}, but
low-luminosity early-type galaxies have $n$ values as high as $\sim
4.5$. NCs in early-type galaxies, like the ones observed by
\cite{Cote+06} are red and old. Since elliptical and S0 galaxies are
thought to have assembled a significant fraction of their mass via
galaxy mergers, we speculate that NCs in early-type galaxies may not
have formed in their present hosts, but in their progenitors, which
may very well have had a less massive spheroidal component and a lower
S\'ersic index $n$ \citep[e.g.,][]{Aguerri+01, NaabTrujillo06}.  This
is also supported by recent cosmological merger simulations, in which
today's elliptical galaxies had to form early-on through the mergers
of gas-rich spiral galaxies to explain the tilt in the Fundamental
Plane \citep[e.g.,][]{Robertson+06}.
% Secular evolution could also have played a role in reshaping the
% NCs' hosts.

Around 20-30\% of spirals and early-type galaxies do not show any
evidence of an NC \citep{Boeker+02}. Especially in spiral galaxies,
this is unlikely due to a lack of gas fuel, but the presence of an
already existing central (dark) mass might suppress the radial tidal
compressive forces throughout the central region.  Remarkably, spirals
without observed NCs tend to show a flattening of the surface
brightness profiles towards the centre, reminiscent of the cores
observed in giant ellipticals.
%% Moreover, the active early-type spirals that are radio-loud
%% show core-like profiles, whereas radio-quiet tend to have power-law
%% profiles \citep[see e.g.,][]{Capetti+07}.

In addition to the surface brightness of galaxies, \cite{Merritt+06}
show that the (deprojected) S\'ersic profile also provides a very good
fit to their (simulated) dark matter halos with $n \sim 3.0$, as well
as those at the scales of clusters with $n \sim 2.4$.  Apart from the
(very) center, we thus expect nearly everywhere disruptive tidal
forces which work against (efficient) formation of stars from
collapsing gas. Whereas in the outer parts of early-type galaxies
there is hardly any gas available, in the intracluster medium of
clusters the tidal forces might contribute in preventing the gas from
cooling. Finally, we should note that the existence of NCs in galaxies
where dark matter is expected to dominate in the central regions
(e.g., some dwarf galaxies) may imply that their corresponding dark
matter halos cannot exhibit very cuspy central profiles, or that
another mechanism prevailed during their formation (if the NCs formed
in situ).

%============================= section 4 =============================
\section{Conclusions}
\label{sec:concl}
%=====================================================================

We have built simple spherical models following deprojected S\'ersic
profiles to show that compressive tidal forces are naturally present
in the central region when the S\'ersic index $n \la 3.5$. For $n \ga
3.5$, the radial component of the tidal forces is disruptive almost
everywhere (i.e., for $r / R_e > 10^{-3}$). Observed nuclear (star)
clusters in early-type and late-type galaxies have extents and/or
apparent locations which are within the tidally compressed region.

If we assume that most NCs form when their host galaxies have density
profiles corresponding to rather low S\'ersic indices $n \sim 1$, we
have shown that the masses of the NCs are consistent with $M_+$, the
mass above which the radial tidal force becomes disruptive due the
presence of the central massive object. In this picture, the predicted
$M_+$ scales linearly with the host galaxy mass (or the mass of the
spheroidal component) with $M_+/\Msph \sim 0.1-0.5$\,\% for $n \sim
1$, in agreement with what is observed for both NCs and super-massive
black holes in the centers of (more luminous) galaxies.  If indeed
compressive tidal forces are a key actor in the formation of CMOs,
only late-type galaxies have, today, the required gas content and
density profiles ($n \sim 1$), which allow the recurrent and common
formation of CMOs (in the form of NCs). This is consistent with the
fact that young NCs are predominantly found in late-type spiral
galaxies. Finally, while we find that tidal compression possibly
drives the formation of CMOs in galaxies, beyond the central regions
and on larger scales in clusters disruptive tidal forces might
contribute to prevent gas from cooling.

Such a simple scenario must be tested and extended to accomodate
galaxies with e.g., core-Sersic surface brightness profiles \citep[see
e.g.][]{Ferrarese+06b} as well as to allow more realistic
(non-spherical, multi-component) galaxy models. Moreover, using
specific (stellar) mass-to-light ratios for the NCs and virial mass
estimates for the host galaxy enables a direct comparison in mass
instead of luminosity. Finally, hydrodynamical simulations are needed
to examine the role of compressive tidal forces in the evolution of
galaxies (and cluster).

%=====================================================================
% ACKNOWLEDGMENTS
%=====================================================================

\section*{Acknowledgments}
\label{sec:acknowledgments}

We thank an anonymous referee for a constructive report which
helped clarifying the text. We would like to sincerely thank Jes\'us
Falc\'on--Barroso, Torsten B\"oker, Pierre Ferruit, Alister Graham,
Richard McDermid, Emmanuel P\'econtal and Scott Tremaine for useful discussions. GvdV
acknowledges support provided by NASA through Hubble Fellowship grant
HST-HF-01202.01-A awarded by the Space Telescope Science Institute,
which is operated by the Association of Universities for Research in
Astronomy, Inc., for NASA, under contract NAS 5-26555.

%=====================================================================
% REFERENCES
%=====================================================================

%\input{ms.bbl}
\bibliographystyle{apj} 
\bibliography{ref_cmo}

%=====================================================================
% APPENDICES
%=====================================================================

\appendix

%============================= appendix A ============================
\section{Deprojected S\'ersic profiles}
\label{app:Abel}
%=====================================================================

The deprojection of the S\'ersic profile (assuming spherical symmetry)
can be solved through an Abel integral equation. This yields for the
(intrinsic) density
\begin{equation}
  \label{eq:densabel}
  \rho(r) = \Upsilon \, \frac{I_e \exp(b_n) b_n}{n \pi R_e}
  \int_r^\infty \left(\frac{R}{R_e}\right)^{1/n-1} 
  \exp\left[-b_n\left(\frac{R}{R_e}\right)^{1/n}\right]
  \frac{\du R}{\sqrt{R^2-r^2}},
\end{equation}
where $\Upsilon$ is the (constant) mass-to-light ratio. Substituting $R =
r \cosh^nu$ gives
\begin{equation}
  \label{eq:densabelu}
  \rho(r) = \Upsilon \, \frac{I_e \exp(b_n) b_n}{n \pi R_e} \;
  \left(\frac{r}{R_e}\right)^{1/n-1} 
  \int_0^\infty \exp\left[-\beta\cosh u\right] \, n \,
  \left(\frac{\cosh^2u-1}{\cosh^{2n}u-1}\right)^{1/2} \du u,
\end{equation}
where $\beta \equiv b_n (r/R_e)^{1/n}$. In general, this integral has
to be solved numerically. 

However, for $n=1$ it reduces to $K_0(\beta)$, with $K_\nu(\beta) =
\int_0^\infty \exp\left[-\beta\cosh u\right] \cosh(\nu u) \, \du u$
the $\nu$\,th-order modified Bessel function of the third kind. For
$n=1/2$, using $\cosh u + 1 = 2\cosh^2(u/2)$, the integral becomes
$K_{1/2}(\beta)/\sqrt{2}$. For other values of $n$, the density is
very well approximated by
\begin{equation}
  \label{eq:densabelapprox}
  \rho(r) \approx  \Upsilon \, \frac{I_e \exp(b_n) b_n}{n \pi R_e} \, 
  \frac{(r/R_e)^{k(1-n)/n} \, 2^{(n-1)/2n} \, K_\nu(\beta)}
  {1- \Sigma_{i=0}^m a_i \log(r/R_e)^i},
\end{equation}
with fitting parameters $\nu$, $k$ and coefficients $a_i$
($i=0,1,\dots,m$) of the $m$\,th-order polynomial in $\log(r/R_e)$.
\cite{Trujillo+02} showed that for $n>1$ a parabolic polynomial
($m=2$) already provides relative errors less than 0.1\,\% in the
radial range $10^{-3} \le r/R_e \le 10^3$. For $0.5<n<1$ (and also
$n<0.5$), a cubic polynomial ($m=3$) is needed to obtain a similarly
good fit. The corresponding best-fit parameters for a range of profiles 
with $0.5 < n < 10$ are provided in Table~\ref{tab:Abelparam}. 

%%%TAB
\begin{table}
\caption{Best-fit parameters for the deprojected S\'ersic profile (Eq.~\ref{eq:densabelapprox}).}
\label{tab:Abelparam}
\begin{tabular}{rccccccc}
\hline
\hline
$n$ & $\nu$ & $k$ & $a_0$ & $a_1$ & $a_2$ & $a_3$ & $\delta_{max}$ \\
(1) & (2) & (3) & (4) & (5) & (6) & (7) & (8) \\
\hline
 0.5 &  0.50000 &  1.00000 &  0.00000 &  0.00000 &  0.00000 &  0.00000 &  0.00000 \\
 0.6 &  0.47768 &  0.85417 & -0.03567 &  0.26899 & -0.09016 &  0.03993 &  0.17568 \\
 0.7 &  0.44879 &  0.94685 & -0.04808 &  0.10571 & -0.06893 &  0.03363 &  0.16713 \\
 0.8 &  0.39831 &  1.04467 & -0.04315 &  0.01763 & -0.04971 &  0.02216 &  0.11766 \\
 0.9 &  0.25858 &  2.55052 & -0.01879 & -0.39382 & -0.08828 & -0.00797 &  0.04783 \\
 1.0 &  0.00000 &  0.00000 &  0.00000 &  0.00000 &  0.00000 &  0.00000 &  0.00000 \\
 1.1 &  0.15502 &  1.59086 &  0.00041 &  0.15211 & -0.03341 &  0.00899 &  0.07371 \\
 1.2 &  0.25699 &  1.00670 &  0.00069 &  0.05665 & -0.03964 &  0.01172 &  0.07741 \\
 1.3 &  0.30896 &  0.88866 &  0.00639 &  0.00933 & -0.04456 &  0.01150 &  0.06961 \\
 1.4 &  0.35245 &  0.83763 &  0.01405 & -0.02791 & -0.04775 &  0.01026 &  0.05948 \\
 1.5 &  0.39119 &  0.81030 &  0.02294 & -0.05876 & -0.04984 &  0.00860 &  0.04964 \\
 2.0 &  0.51822 &  0.76108 &  0.07814 & -0.16720 & -0.05381 & -0.00000 &  0.02943 \\
 2.5 &  0.53678 &  0.83093 &  0.13994 & -0.13033 & -0.03570 & -0.00000 &  0.02576 \\
 3.0 &  0.54984 &  0.86863 &  0.19278 & -0.10455 & -0.02476 &  0.00000 &  0.01790 \\
 3.5 &  0.55847 &  0.89233 &  0.23793 & -0.08618 & -0.01789 & -0.00000 &  0.01233 \\
 4.0 &  0.56395 &  0.90909 &  0.27678 & -0.07208 & -0.01333 &  0.00000 &  0.00865 \\
 4.5 &  0.57054 &  0.92097 &  0.31039 & -0.06179 & -0.01028 & -0.00000 &  0.00587 \\
 5.0 &  0.57950 &  0.93007 &  0.33974 & -0.05369 & -0.00812 & -0.00000 &  0.00386 \\
 5.5 &  0.58402 &  0.93735 &  0.36585 & -0.04715 & -0.00653 & -0.00000 &  0.00277 \\
 6.0 &  0.58765 &  0.94332 &  0.38917 & -0.04176 & -0.00534 & -0.00000 &  0.00203 \\
 6.5 &  0.59512 &  0.94813 &  0.41003 & -0.03742 & -0.00444 & -0.00000 &  0.00145 \\
 7.0 &  0.60214 &  0.95193 &  0.42891 & -0.03408 & -0.00376 & -0.00000 &  0.00105 \\
 7.5 &  0.60469 &  0.95557 &  0.44621 & -0.03081 & -0.00319 & -0.00000 &  0.00082 \\
 8.0 &  0.61143 &  0.95864 &  0.46195 & -0.02808 & -0.00274 & -0.00000 &  0.00061 \\
 8.5 &  0.61789 &  0.96107 &  0.47644 & -0.02599 & -0.00238 & -0.00000 &  0.00047 \\
 9.0 &  0.62443 &  0.96360 &  0.48982 & -0.02375 & -0.00207 & -0.00000 &  0.00036 \\
 9.5 &  0.63097 &  0.96570 &  0.50223 & -0.02194 & -0.00182 & -0.00000 &  0.00028 \\
10.0 &  0.63694 &  0.96788 &  0.51379 & -0.02004 & -0.00160 &  0.00000 &  0.00022 \\
\hline
\end{tabular}
\tablecomments{Notes:
(1)~S\'ersic index. 
(2)--(7)~Best-fit parameters. 
(8)~Relative maximum error (in \%)}
\end{table}
%%%TAB

%=====================================================================
% END DOCUMENT
%=====================================================================

\end{document}